\documentclass[plb%
 ,twocolumn
,secnumarabic%
,tightenlines%
,aps
,floatfix
,nofootinbib
]{revtex4}

\usepackage{graphicx}
\usepackage{color}
\usepackage{amssymb}
\usepackage{amsmath}
\usepackage{docs}%
\usepackage{bm}%
\usepackage[colorlinks=true,linkcolor=blue]{hyperref}

\newcommand{\vev}[1]{{\langle #1 \rangle}}

\newcommand{\MeV}{\mbox{~MeV}}
\newcommand{\GeV}{\mbox{~GeV}}

\newcommand{\ie}{{\it i.e.}}
\newcommand{\eqn}[1]{&\hspace{-0.1em}#1\hspace{-0.1em}&}

\newcommand{\simgeq}{\mbox{$\hspace{0.3em}\raisebox{0.4ex}{$>$}\hspace{-0.75em}\raisebox{-.7ex}{$\sim$}\hspace{0.3em}$}}

\begin{document}
\title{\boldmath{
Realisability of SUSY discrete flavour symmetry in our universe.
}}
\author{Hiroaki~Nagao$^{1}$ and Yusuke~Shimizu$^{2}$}
\affiliation{$^{1}$Graduate School of Science and Technology, Niigata University, 950-2181, Japan\\ 
$^{2}$Department of Physics, Niigata University, 950-2181, Japan}
\date{\today}
\begin{abstract}{
    We study the realisation of supersymmetric discrete flavour symmetry models to the thermal history
    of our universe.
    We focus on the evolution of the pseudo moduli field among the flavons
    by taking into account finite temperature corrections.
    We show that the pseudo moduli flavon dominates the energy density of our universe and 
    this domination makes crucially difficult to realise the flavour symmetry models in our universe. 
    We also discuss possible extensions of the supersymmetric discrete flavour symmetry models
    which can ensure the consistency of the models with the thermal history of our universe.
    Finally, we show an extension to realise the thermal inflation by the flavon domination.
 }   
\end{abstract}
\maketitle
%%%%%%%%%%%%%%%%%%%%%%%%%%%%%%%%%%%%%%%%%%%%%%%%%%%%%%%%%%%%%%%%%%%%
\section{Introduction}
\label{sec:intro}
Neutrino experimental data provides valuable information on the origin of the observed hierarchies in the lepton masses. 
Recent experiments of the neutrino oscillation indicate the almost tri-bimaximal mixing~\cite{Harrison:2002er} for three flavours in the lepton sector,
except for large $\theta_{\rm 13}$~\cite{LargeTH13}.
On theoretical aside, many flavour symmetry models have been proposed 
in order to explain the lepton masses and mixing angles so far.
In particular, non-Abelian discrete symmetry, like $A_4$ or $S_4$, have achieved great success
in deriving the tri-bimaximal mixing~\cite{Ishimori:2010au}.

In the discrete flavour symmetry models, there are a large number of scalar fields, so-called {\it flavon}.
The vacuum expectation values (VEVs) of flavons can provide the flavour structures of the standard model (SM) fermions
through the {\it vacuum alignment}~\cite{Altarelli:2005yx}.
It has been widely argued that various cosmological issues can be explained
by the degrees of freedom of flavon fields, for instance, the leptogenesis scenario with low reheating temperature~\cite{Riva:2010jm} or
realisation of inflationary universe~\cite{Antusch:2008gw}. 

Especially in the supersymmetric (SUSY) discrete flavour symmetry models, there exists a F-flat direction among the flavons.
It is generally expected that the flavour symmetry is spontaneously broken along the flat direction among flavons (or equivalently the pseudo moduli flavon)
at sufficiently high temperature in the history of our universe.

In this letter, we study the evolution of the pseudo moduli flavon in the SUSY discrete flavour symmetry models without relying on specific models.
We show that the potential energy of the pseudo moduli flavon dominates the universe at high temperatures,
and such domination conflicts with successes of the standard Big Bang cosmology.
Then, we propose a non-trivial extension of the models to overcome the problem in the SUSY discrete flavour symmetry.
With this extension, the domination by the flavon potential energy can sufficiently dilute dangerous relics of the cosmological modulus.
Hereafter, we denote the pseudo moduli flavon and its VEV as $u$ and $\vev{u}$, respectively.

\section{Problem of pseudo moduli flavon}
\label{sec:problem}
In a large class of the SUSY discrete flavour symmetry models, the pseudo moduli flavon is coupled with right-handed neutrinos $\nu_R$
and the VEV of $u$ provides the majorana mass term of $\nu_R$'s which is expected to be around $10^{14}\GeV$
to ensure the See-Saw mechanism with order one yukawa couplings~\cite{Seesaw}.
We shall consider this class of discrete flavour symmetry models below.

We first give an  analytic expression for 
the VEV of $u$ at zero temperature ($T=0$)
by including SUSY breaking effects and 1-loop effective potentials. 
Next we shall discuss the problem of pseudo moduli flavon in the history of our universe 
by taking into account finite temperature corrections.
\subsection{Evolution of flavon at $T=0$}
\label{sec:zeroT}
In order to realise the spontaneous symmetry breaking, the flavon potential should have a negative curvature around its origin.
In the SUSY discrete flavour symmetry models at low energies,
there is a strong candidate for such negative curvature; namely, a tachyonic soft SUSY breaking mass of $u$.

The 1-loop effective potentials induced by the right-handed (s)neutrinos $\nu_R$ ($\tilde{\nu}_R$) lift the flavon potential at large field value of $u$.
The potential energy of $u$ including the soft SUSY breaking mass $\tilde{m}_u$ and 1-loop effective potentials at zero temperature
can be written by $V(u) = V_0 - \tilde{m}_u^2 |u|^2+\Delta V_{\nu_R} +\Delta V_{\tilde{\nu}_R} $, 
where $V_0$ is the potential energy at $u=\vev{u}$ which is chosen as $V(\vev{u})=0$.
$\Delta V_{\nu_R} $ and $\Delta V_{\tilde{\nu}_R} $ are the 1-loop effective potentials from the coupling with $\nu_R$ and $\tilde{\nu}_R$ respectively, 
\begin{eqnarray}
\label{eq:CWpot_RHnu}
\Delta V_{\nu_R} \eqn{=}
	-\sum_{i=1}^3 \frac{M_{\nu_R i}(u)^4}{32 \, \pi^2}
		\left[ \ln{ \left( \frac{  M_{\nu_R i}(u)^2  }{ \mu^2 } \right) }-\frac{3}{2}\right]\, ,\nonumber\\
\Delta V_{\tilde{\nu}_R} \eqn{=}
	\sum_{i=1}^3 \frac{M_{\tilde{\nu}_R i}(u)^4}{32 \, \pi^2}
		\left[ \ln{ \left( \frac{ M_{\tilde{\nu}_R i}(u)^2  }{ \mu^2 } \right) }-\frac{3}{2}\right]\, ,
\end{eqnarray}
where $M_{\nu_R i}(u)^2\equiv |y_{R i}|^2 |u|^2$, $M_{\tilde{\nu}_R i}(u)^2\equiv |y_{R i}|^2 |u|^2+\tilde{m}_{\nu_Ri }$ \cite{Bando:1993ma},
and the index $i$ represents the generation (flavour) of right-handed neutrinos.
Here, $y_{R\, i}$ is yukawa coupling between $u$ and $\nu_{R \scriptstyle i}$, $\tilde{m}_{\nu_R\, i}^2$ is soft SUSY-breaking mass of $\tilde{\nu}_ {R \scriptstyle i}$, 
and $\mu$ is energy scale where we evaluate the potential.
\footnote{In our analysis, we will take $\mu = \tilde{m}_u$.}
We take both $y_{R\,i}$'s and $\tilde{m}_{\nu_R\, i}^2$'s to be the same order respectively for simplicity.
Note that the A-terms do not contribute to $V(u)$ because they are forbidden by $U(1)_R$ symmetry~\cite{Altarelli:2005yx}.
We stress that any soft SUSY breaking terms do not lift $V(u)$ at large field value of $u$,
even if the higher dimensional operators are taken into account, in contrast to the analysis in~\cite{Riva:2010jm}.
\footnote{The author in~\cite{Riva:2010jm} claimed that the SUSY breaking term, $\Lambda_{\rm EW}^2 (u^3+u^{\ast 3})/\Lambda$, lifts the flavon potential
at large field value of $u$.
However, such a term is never induced once all of the next leading order terms of $1/\Lambda$ is consistently taken into account.
The relevant terms does not appear due to the definition of F-flat direction by including the next leading terms.}

By taking the limit of large field value in Eq.(\ref{eq:CWpot_RHnu}),
\ie, $|u| \gg \mu$, the flavon potential is reduced  to
\begin{eqnarray}
V(u) \simeq \left[-\tilde{m}_u^2 
	+ \frac{3 \, |y_R|^2 \, \tilde{m}_{\nu_R}^2 }{16 \, \pi^2}\ln{ \left( \frac{  |y_R|^2 \, |u|^2  }{ \tilde{m}_u^2 } \right) }
	 \right]\, |u|^2  .
\end{eqnarray}
Here, we replace the yukawa couplings $y_{R\, i}$ by the averaged value $y_R$,
the validity of which will be discussed elsewhere \cite{Nagao:2012fw}.
Consequently, we obtain the analytic expression for $\vev{u}$,
\begin{equation}
\label{eq:vev_approx1}
\langle u \rangle 
 \simeq \frac{\tilde{m}_u}{\sqrt{|y_R|^2}} \exp \left( \frac{8\pi^2 \tilde{m}_u^2}{3 \, |y_R|^2 \, \tilde{m}_{\nu_R}^2} \right) \, .
 \end{equation}
We see that the VEV of the pseudo moduli flavon strongly depends on the soft SUSY breaking masses and the yukawa couplings.
We also obtain analytical expressions for the mass of $u$ around the vacuum and $V_0$ as
\begin{eqnarray}
m_u^2 \simeq\frac{3 \, |y_R|^2 }{4 \, \pi^2 }\, \tilde{m}_{\nu_R}^2\, , ~~ V_0\simeq\frac{3 \, |y_R|^2}{16 \, \pi^2 } \, \tilde{m}_{\nu_R}^2 \,\langle u \rangle^2 \, ,
\end{eqnarray}
respectively.
 \subsection{Thermal evolution of flavon at $T\ne 0 $}
 \label{sec:nonzero}
Once we take into account the finite temperature effect,
the quadratic thermal potential $T^2 |u|^2$ contributes to $V(u)$ at high cosmic temperatures. 
The pseudo moduli flavon $u$ is trapped at the origin due to this positive curvature.
Although, one might expect that the discrete flavour symmetry is spontaneously broken at sufficiently high temperatures,
such symmetry breaking is never simply realised due to the thermal corrections.

In the epoch of high cosmic temperatures,
the potential energy is kept to almost constant until $u$ rolls down to its minimum, \ie,
\begin{eqnarray}
\label{eq:Pot-u_naive}
V(u,T) =V(u,T=0)+c\, T^2 |u|^2\simeq V_0 \quad ( T \gg \tilde{m}_u )\, ,
\end{eqnarray}
where the factor $c$ depends on coupling constants in the models and is taken to be ${\cal O}(1)$ in our following analysis.

As long as the cosmic temperature satisfies  $V_0^{1/4} >T > \tilde{m}_u$, the flavon potential energy $V_0$ dominates the energy density of our universe.
We find that the flavon domination occurs in many classes of models~\cite{Ding:2011qt,Hagedorn:2010th,Adulpravitchai:2008yp}, where a F-flat direction among flavons exists 
and the VEV of the flat direction is expected to be larger than $\Lambda_{\rm EW}$.
We will show that achievements of the standard cosmology are ruined in consequence of the flavon domination.

When the cosmic temperature decreases below $\tilde{m}_u$, 
the pseudo moduli flavon $u$ rolls down to the minimum of the potential and obtains a non-vanishing VEV,
$\vev{u} \neq 0$.
The energy of $u$ is simultaneously stored into the SM particles through the decays of $u$.

In order to recover the standard Big Bang cosmology after the flavon domination,  the decay of $u$ into the SM particles must be completed
before the Big Bang nucleosynthesis (BBN).
The temperature $T_R^{(sec)}$ of the energy transfer is estimated as $T_R^{(sec)}\simeq 0.5 \sqrt{\Gamma(u\hspace{-1.5pt}\rightarrow \hspace{-1.5pt}{\rm SM}) M_P}$,
where $M_p$ is the reduced Planck mass.
We have to build the model to satisfy the condition $T_R^{(sec)} \simgeq {{\cal O}(1) \,{\rm MeV}}$ not to conflict with BBN achievements.

In order to verify whether $T_R^{(sec)}$ can be above a few MeV, we estimate the decay rate of $u$.
The interactions of $u$ in our analysis are given by
\begin{equation}
\label{eq:Lint_naive}
{\cal L}_{int} \ni \frac{y_R}{\sqrt{2}}\delta u \, \nu^c_R \nu_R^c+h.c. \, + \,  \frac{\delta u}{\sqrt{2}\, \vev{u}}\left(\partial_\mu a\right)^2\, .
\end{equation}
Here we decompose $u$ as
\begin{equation}
u \equiv \left( \vev{u}+\frac{\delta u}{\sqrt{2}}\right)\exp{\left(i \frac{a}{\sqrt{2}\vev{u}}\right)}\, ,
\label{eq:u_decomp}
\end{equation}
where $a$ denotes the NG boson associated with the spontaneous breaking of $U(1)$ symmetry in $V(u)$.
\footnote{In the $A_4$ or $S_4$ symmetry model, this $U(1)$ symmetry can be interpreted to the global $U(1)_{B-L}$ symmetry which is implicitly included 
in the superpotential. }
Using the interactions in Eq.(\ref{eq:Lint_naive}), the most efficient decay process of $u$ into the SM particles is found to be
 $u \hspace{-1.5pt}\rightarrow  \hspace{-1.5pt}2\nu$, and the decay rate is given by
\begin{eqnarray}
\Gamma(u \hspace{-1.5pt}\rightarrow  \hspace{-1.5pt}2 \nu)\eqn{\simeq} \frac{3}{16\, \pi}\left(\frac{m_\nu}{\vev{u}}\right)^2 m_u\, .
\end{eqnarray}

On the other hand, $u$ also decays into the NG boson $a$ through the derivative coupling.
It is easy to find that the decay process $u \hspace{-1.5pt}\rightarrow  \hspace{-1.5pt} 2 a$ dominates the decays of $u$,
and the decay rate is given by
\begin{eqnarray}
\Gamma(u  \hspace{-1.5pt}\rightarrow  \hspace{-1.5pt}2a)\eqn{\simeq}\frac{1}{64\pi}\frac{m_u^3}{\vev{u}^2}\, .
\end{eqnarray}
Hence, the most of the energy of $u$, which dominates the energy of our universe, will be stored into the NG boson $a$.
Once the energy of $u$ is released into $a$, 
the energy of $a$ is never released into the SM particles before the BBN begins,
because the interaction rate of $a$ is much smaller than one corresponding to $T_R^{(sec)}\sim {\rm few \,MeV}$.

According to general discussion given above, 
we conclude that the SUSY discrete flavour symmetry models, for example the models given in~\cite{Ding:2011qt,Hagedorn:2010th,Adulpravitchai:2008yp},
do not agree with the thermal history of our universe,
since the energy of our universe is overclosed by the NG boson $a$.

\section{ Possible improvements}
\label{sec:improve}
The essential point of the problem of the overclosure by $a$ attributes to the largeness of Br$(u \hspace{-1.5pt}\rightarrow \hspace{-1.5pt} 2a)$
and the smallness of the interaction rate of the NG boson $a$.
So long as the discrete flavour symmetry is retained rigorously,
we should extend the models to increase Br$(u \hspace{-1.5pt}\rightarrow \hspace{-1.5pt} {\rm SM})$,
since the decay of $u$ into massless NG boson $a$ is always opened.
\subsection{Introduction of $uQ \overline{Q}$ coupling}
\label{sec:uQQ}
First we consider the extension by introducing new interaction into the superpotential as
$\Delta W = y_Q\, u \, Q \overline{Q}\, $.
Here, $Q$ and $\overline{Q}$ are additional vector-like ``heavy quarks" which have $SU(3)_{\rm QCD}$ charge.
With the aid of this coupling, the pseudo moduli flavon $u$ can decay into gluons through the ``heavy quark'' loops.
The decay rate of $u \hspace{-1.5pt}\rightarrow \hspace{-1.5pt} g g$ is given by \cite{Chun:2000jr}
\begin{eqnarray}
\Gamma(u \hspace{-1.5pt}\rightarrow \hspace{-1.5pt} gg)=\frac{N_q^2\alpha_s^2}{144\pi^3} \frac{m_u^3}{\vev{u}^2}\, , 
\label{eq:Gamma_2g}
\end{eqnarray}
where $N_q$ denotes the number of flavours of $Q,\overline{Q}$.

The branching ratio of $u \hspace{-1.5pt}\rightarrow \hspace{-1.5pt}{\rm SM}$ is now found to be
\begin{equation}
{\rm Br} (u \hspace{-1.5pt}\rightarrow \hspace{-1.5pt} {\rm SM})
\simeq \frac{\Gamma(u \hspace{-1.5pt}\rightarrow \hspace{-1.5pt} gg)}{\Gamma(u \hspace{-1.5pt}\rightarrow \hspace{-1.5pt} 2a)}=\frac{4N_q^2 \alpha_s^2}{9\pi^2}\, .
\end{equation}
To overcome the decay into $a$, about a hundred  ``heavy quarks'' should be introduced.
According to the discussion in \cite{Morrissey:2005uz}, however, the running gauge coupling blows up
due to the corrections from a large number of ``heavy quarks'' below the GUT scale.
Thus, this kind of modification is not favoured.
\subsection{Light right-handed neutrinos}
\label{sec:lightRH}
Since the process $u \hspace{-1.5pt}\rightarrow \hspace{-1.5pt}2\nu$ contains the propagators of heavy $\nu_R$,
the decay rate of $u \hspace{-1.5pt}\rightarrow \hspace{-1.5pt}{\rm SM}$ is suppressed by $1/\vev{u}^2$. 
If the pseudo moduli flavon can decay directly into the right-handed neutrinos,
the branching ratio of $u \hspace{-1.5pt}\rightarrow \hspace{-1.5pt}{\rm SM}$ will be enhanced,
because the right-handed neutrinos produced by a decay of $u$ subsequently decay into active neutrinos and higgses.
Thus, one might naively expect that the model with relatively light right-handed neutrinos
\footnote{The masses of light right-handed neutrinos are induced from the higher dimensional operator, \ie,  $W\ni y_R u^2 \nu_R^c\nu_R^c/\Lambda$~\cite{Nagao:2012fw}}
could be a candidate for the modification of the model to overcome the problem of the NG boson $a$.
With this extension, the interactions of $u$ are given by ${\cal L}_{int} = \frac{y_R}{\sqrt{2}}\frac{\vev{u} }{\Lambda}\, \delta u\, \nu^c_R \nu_R^c+h.c.$
and the branching ratio of the process $u \hspace{-1.5pt}\rightarrow  \hspace{-1.5pt}2a$ is given by 
\begin{eqnarray}
\hspace{-12pt}{\rm Br}(u \hspace{-1.5pt}\rightarrow \hspace{-1.5pt} 2a)\simeq\left[1+24 \left(\frac{m_{\nu_R}}{m_u}\right)^{\hspace{-3pt}2}\left(1-\frac{4m_{\nu_R}^2}{m_u^2}\right)^{\hspace{-3pt}3/2}\right]^{-1}\hspace{-3pt}.
\end{eqnarray}
According to Eq.(12), Br$(u \hspace{-1.5pt}\rightarrow \hspace{-1.5pt} {\rm SM})$ never exceed 50 percent,
even if the right-handed neutrinos are light enough.
When Br$(u \hspace{-1.5pt}\rightarrow \hspace{-1.5pt}2a)$ is not less than 45 percent,
the relativistic NG boson $a$ significantly modifies the abundance of light elements produced in BBN due to the ``speed-up effect"~\cite{Cyburt:2004yc,Komatsu:2010fb}.
Thus, this kind of modification is not efficient for the problem of overclosure by the NG boson.
\subsection{Explicit breaking term of flavour symmetry}
\label{sec:ExpBre}
Even if we introduce the new fields which have QCD charge, or  the sufficiently light right-handed neutrinos, 
the branching ratio of Br$(u \hspace{-1.5pt}\rightarrow  \hspace{-1.5pt}{\rm SM})$ does not mainly contribute to the total decay width of $u$.
Hence, we consider an extension of the model, which kinematically forbid the process $u \hspace{-1.5pt}\rightarrow \hspace{-1.5pt} 2a$.
In order to give the mass to the NG boson $a$,
we assume an explicit breaking term for the $U(1)$ symmetry in the flavon potential.
\footnote{Here, we do not specify the origin of the explicit breaking term.}
This also breaks discrete flavour symmetry explicitly, as
\begin{equation}
\Delta {\cal L}= -\kappa \left(u+u^\ast\right)\, ,
\end{equation}
where $\kappa$ is a dimensionful parameter.
This explicit breaking term gives the NG boson mass $m_a^2=\kappa/\langle u \rangle$.
Again, we stress that such explicit breaking term cannot be induced by the SUSY breaking effects or higher dimensional operators.

With the help of the explicit breaking term, the decay process $u \hspace{-1.5pt}\rightarrow  \hspace{-1.5pt}2a$ is kinematically forbidden,
when $\kappa > m_u^2 \vev{u}/4 $. 
To ensure the sizeable $T_R^{(sec)}$, we further introduce the interaction with vector-like ``heavy quarks'', $\Delta W = y_Q u Q \overline{Q}$,
with a small number of flavours.
\footnote{ The number of flavour of ``quarks" should be less than eight for {\bf 5} representation of $SU(5)$~\cite{Morrissey:2005uz}.}
The energy transfer from the flavon occurs by the process $u \hspace{-1.5pt}\rightarrow  \hspace{-1.5pt}g g$ which can make $T_R^{(sec)}$ around MeV and 
the width of the process is given by Eq.(\ref{eq:Gamma_2g}).
We find  that this extension is only the possible solution for the problem of the overclosure by the NG boson after the flavon domination.

\section{Realisation of thermal inflation}
\label{sec:flavonTI}
As we have already explained, the pseudo moduli flavon $u$ is trapped at the origin for $T \simgeq T_{\rm end} =\tilde{m}_u$, and
the flavon domination occurs when the cosmic temperature decreases to $T\simeq V_0^{1/4}$.
If we consider the model including the explicit breaking term and the coupling with the vector-like ``heavy quarks'',
the pseudo moduli flavon $u$ mainly decays into gluons.
It allows us to apply the flavon domination to the thermal inflation scenario \cite{Barreiro:1996dx},
which is the mechanism for diluting the abundance of dangerous cosmological modulus $\chi$.

In the presence of thermal inflation, the energy density of cosmological modulus is diluted 
through the entropy production after thermal inflation. 
The modulus abundance after the thermal inflation is given by \cite{Asaka:1999xd}
\begin{eqnarray}
\frac{\rho_\chi}{s}(T^{(sec)}_R)=\frac{1}{\Delta}\frac{1}{8}T_R \left(\frac{\chi_0}{M_{P}}\right)^2\, ,
\end{eqnarray}
where $\Delta$ is entropy dilution factor which is defined by $\Delta \equiv \frac{30}{\pi^2 g_*(T_{\rm end})}\frac{V_0}{T_{\rm end}^3 T^{(sec)}_R}$,
$\chi_0$ the initial amplitude of cosmological modulus, and $T_R$ the reheating temperature after primordial inflation.

The relevant quantities are estimated as
\begin{eqnarray}
\label{eq:vev_approx2}
\eqn{}\vev{u}\simeq\frac{\tilde{m}_u}{\sqrt{|y_{\rm ave}|^2}}\exp{\left[\frac{8\pi^2\tilde{m}_{u}^2}{(3+2 N_q)|y_{\rm ave}|^2\tilde{m}_{\rm ave}^2}\right]}\nonumber\\
\eqn{}\qquad\simeq1\times 10^{13} \GeV \, ,\\
\eqn{}T_R^{(sec)}=\left(\frac{90}{\pi^2 g_\ast}\right)^{1/4}\sqrt{\Gamma(u\rightarrow gg) M_P}\simeq 2\MeV \, , \\
\eqn{}\left (\frac{\rho_\chi}{s}\right)
         \simeq 7\times 10^{-19}{\rm GeV}\left(\frac{T_R}{10^6 {\rm GeV}}\right)\, .\label{eq:abundance_mod}
\end{eqnarray}
where $|y_{\rm ave}|$ and $\tilde{m}_{\rm ave}$ denote the averaged values of $y_R , y_Q$ and $\tilde{m}_{Q},\tilde{m}_{\bar{Q}},\tilde{m}_{\nu_R}$ respectively.
Here, we set $\kappa =m_u^2 \vev{u}/4$,  $N_q=8$, $|y_{\rm ave}|=1  ,\, \tilde{m}_u=500\GeV $, and $\tilde{m}_{\rm ave}=100\GeV$.
Eq.(\ref{eq:abundance_mod}) indicates that the modulus abundance can be sufficiently diluted by thermal inflation 
caused by the pseudo moduli flavon $u$.
The details of the realisability of thermal inflation by $u$ are shown elsewhere~\cite{Nagao:2012fw}.

%%%%%%%%%%%%%%%%%%%%%%%%%%%%%%%%%%%%%%%%%%%%%%%%%%%%%%%%%%%%%%%%%%%%
%%%%%%%%%%%%%%%%%%%%%%%%%%%%%%%%%%%%%%%%%%%%%%%%%%%%%%%%%%%%%%%%%%%%
%%%%%%%%%%%%%%%%%%%%%%%%%%%%%%%%%%%%%%%%%%%%%%%%%%%%%%%%%%%%%%%%%%%%
\section{Summary and Discussions}
 We studied the thermal evolution of the pseudo moduli flavon, which is included in a large number of SUSY discrete flavour symmetry models.
 The pseudo moduli flavon has the potential energy due to
 the finite temperature corrections, 1-loop effective potentials, and the SUSY breaking masses.
Especially, we considered the model including the coupling between the pseudo moduli flavon and the right-handed neutrinos.

We first gave the analytic expression for the VEV of the pseudo moduli flavon at zero temperature.
The scale of the VEV of the pseudo moduli flavon strongly depends on the SUSY breaking masses and yukawa couplings.

At finite temperature $T>\tilde{m}_u$, 
the pseudo moduli flavon is trapped at the origin due to the thermal corrections from the heat bath. 
During this epoch, the pseudo moduli flavon starts to dominate the energy density of our universe, when the cosmic temperature decreases below $V_0^{1/4}$. 
After the flavon domination ends, the pseudo moduli flavon mostly decays into the NG boson associated with the spontaneous discrete flavour symmetry breaking.
The temperature of energy transfer from $a$ into the SM particles is much lower than  few MeV, since the interaction rate of $a$ is too small.
Hence, the energy of our universe is overclosed by the NG boson $a$ and the realisation of SUSY discrete flavour symmetry models becomes difficult significantly.
 
To overcome the problem of the overclosure by the NG boson, 
we examined the three possible extensions of the SUSY discrete flavour symmetry model :
(i)  Introducing the interaction with vector-like ``heavy quarks'' into the superpotential.
It opens the decay channel into gluons through the ``heavy quark" loops.
Although this extension could throw out the problem with hundred flavours,
such a large number of ``heavy quarks" derive the gauge couplings to blow up below GUT scale.
(ii) The model with light right-handed neutrinos.
Although this can open the decay channel into the right-handed neutrinos of on-shell,
the branching fraction of this process cannot exceed 50 percent.
When the branching ratio of the decay process, (flavon) $\rightarrow$ (SM), is less than 50 percent,
the relativistic NG boson significantly modify the abundance of light nucleus produced in BBN.
(iii) Assuming the explicit breaking term of the SUSY discrete flavour symmetry in the lagrangian.
This breaking term induces the NG boson mass
and the decay of pseudo moduli flavon into the NG boson is forbidden by the sufficiently large breaking term.
However, the origin of the explicit breaking term is unclear,
since such term is never induced by any SUSY breaking effects or the higher dimensional operators,
as long as the flavour symmetry is retained exactly.

We found that the combination of (i) with the small number of flavours of ``heavy quarks" and (iii) 
is the only way to avoid the problem of the overclosure by the NG boson,
and such a combination makes the temperature of the energy transfer to be above a few MeV.

Finally, we showed the applicability of discrete flavour model to the thermal inflation scenario.
The abundance of cosmological modulus is sufficiently diluted by the flavon domination.

We close this letter by pointing out the future tasks on our study.
In this letter, the values of soft SUSY breaking masses are taken as free parameters.
However, if we specify the mediation mechanism for the SUSY breaking, the size of soft SUSY breaking term would be determined by the SUSY breaking scales.
The scale of the VEV of the pseudo moduli flavon is sensitive to the variance of parameters.
Therefore the radiative corrections to these parameters would significantly change the scale of the VEV even if those corrections are small.

We also comment on another possibility to generate the NG boson mass $m_a^2$ in different analysis.
Our analysis in this letter was based on the F-flat direction and unbroken $U(1)_R$ symmetry,
and that leads to a massless NG boson $a$.
Once the  $U(1)_R$ symmetry is broken and the nonrenormalisable K\"{a}hler potential, such as $K\ni (u^3+u^{\dagger 3})/M_p$ for $A_4$ model and
$K\ni (u^4+u^{\dagger 4})/M_p^2$ for $S_4$ model~\cite{Altarelli:2005yx,Ding:2011qt,Hagedorn:2010th}, is introduced,
the NG boson can obtain the mass in analogy with the R-axion~\cite{Nelson:1993nf}.
The mass of NG boson $a$ in this case is roughly estimated as
$m_a^2 \propto m_{3/2} \vev{u}$ and $m_a^2 \propto m_{3/2}\vev{u}^2/M_p$ for $A_4$ and $S_4$ models respectively.
However, within our knowledge, it has not been clarified whether the flavour structure of the SM fermions is altered
by the broken $U(1)_R$ symmetry and the nonrenormalisable K\"{a}hler potential.
Detailed studies will be given elsewhere~\cite{Nagao:2012fw}.

\section*{Acknowledgements}
%%%%%%%%%%%%%%%%%%%%%%%%%%%%%%%%%%%%%%%%%%%%%%%%%%%%%%%%%%%%%%%%%%%%
The work of H.~N and Y.~S were supported by Grant-in-Aid for JSPS Fellows No.23.6174 and No.22.3014 respectively.
We would like to thank K.~Nakayama for helpful discussions in early stage of this work.
We also would like to thank Y.~Nakagawa and T.~Shimomura for helpful discussions during this collaboration. 
%%%%%%%%%%%%%%%%%%%%%%%%%%%%%%%%%%%%%%%%%%%%%%%%%%%%%%%%%%%%%%%%%%%%
%%%%% ** Reference ** %%%%%%%%%%%%%%%%%%%%%%%%%%%%%%%%%%%%%%%%%%%%%%
%%%%%%%%%%%%%%%%%%%%%%%%%%%%%%%%%%%%%%%%%%%%%%%%%%%%%%%%%%%%%%%%%%%%

%%%%%%%%%%%%%%%%%%%%%%%%%%%%%%%%%%%%%%%%%%%%%%%%%%%%%%%%%%%%%%%%%%%%
%%%%%%%%%%%%%%%%%%%%%%%%%%%%%%%%%%%%%%%%%%%%%%%%%%%%%%%%%%%%%%%%%%%%
%%%%%%%%%%%%%%%%%%%%%%%%%%%%%%%%%%%%%%%%%%%%%%%%%%%%%%%%%%%%%%%%%%%%

\begin{thebibliography}{100}
%\cite{Harrison:2002er}
\bibitem{Harrison:2002er} 
  P.~F.~Harrison, D.~H.~Perkins and W.~G.~Scott,
  %``Tri-bimaximal mixing and the neutrino oscillation data,''
  Phys.\ Lett.\ B {\bf 530}, 167 (2002)
  [hep-ph/0202074];
  %%CITATION = HEP-PH/0202074;%%
  P.~F.~Harrison and W.~G.~Scott,
  %``Symmetries and generalizations of tri - bimaximal neutrino mixing,''
  Phys.\ Lett.\ B {\bf 535}, 163 (2002)
  [hep-ph/0203209].
  %%CITATION = HEP-PH/0203209;%%
  \bibitem{LargeTH13}
%\cite{An:2012eh}
  F.~P.~An {\it et al.}  [DAYA-BAY Collaboration],
  %``Observation of electron-antineutrino disappearance at Daya Bay,''
  Phys.\ Rev.\ Lett.\  {\bf 108}, 171803 (2012)
  [arXiv:1203.1669 [hep-ex]];
  %%CITATION = ARXIV:1203.1669;%%;
%\cite{Ahn:2012nd}
  J.~K.~Ahn {\it et al.}  [RENO Collaboration],
  %``Observation of Reactor Electron Antineutrino Disappearance in the RENO Experiment,''
  Phys.\ Rev.\ Lett.\  {\bf 108}, 191802 (2012)
  [arXiv:1204.0626 [hep-ex]].
  %%CITATION = ARXIV:1204.0626;%%
%\cite{Ishimori:2010au}
\bibitem{Ishimori:2010au} 
For example, See  H.~Ishimori, T.~Kobayashi, H.~Ohki, Y.~Shimizu, H.~Okada and M.~Tanimoto,
  %``Non-Abelian Discrete Symmetries in Particle Physics,''
  Prog.\ Theor.\ Phys.\ Suppl.\  {\bf 183}, 1 (2010)
  [arXiv:1003.3552 [hep-th]].
  %%CITATION = ARXIV:1003.3552;%%
  %\cite{Altarelli:2005yx}
\bibitem{Altarelli:2005yx} 
  G.~Altarelli and F.~Feruglio,
  %``Tri-bimaximal neutrino mixing, A(4) and the modular symmetry,''
  Nucl.\ Phys.\ B {\bf 741}, 215 (2006)
  [hep-ph/0512103].
  %%CITATION = HEP-PH/0512103;%%
%\cite{Riva:2010jm}
\bibitem{Riva:2010jm} 
  F.~Riva,
  %``Low-Scale Leptogenesis and the Domain Wall Problem in Models with Discrete Flavor Symmetries,''
  Phys.\ Lett.\ B {\bf 690}, 443 (2010)
  [arXiv:1004.1177 [hep-ph]].
  %%CITATION = ARXIV:1004.1177;%%
%\cite{Antusch:2008gw}
\bibitem{Antusch:2008gw} 
  S.~Antusch, S.~F.~King, M.~Malinsky, L.~Velasco-Sevilla and I.~Zavala,
  %``Flavon Inflation,''
  Phys.\ Lett.\ B {\bf 666}, 176 (2008)
  [arXiv:0805.0325 [hep-ph]].
  %%CITATION = ARXIV:0805.0325;%%
  \bibitem{Seesaw}
 P. Minkowski, Phys. Lett.
{\bf{B67}},110(1977);
 M.~Gell-Mann, P.~Ramond and R.~Slansky,
in {\it Supergravity}, eds. P.~van~Niewenhuizen and D.Z.~ Freedman
(North Holland 1979); T.~Yanagida, in Proceedings of {\it Workshop
on Unified Theory and Baryon number in the Universe}, eds.
O.~Sawada and A. Sugamoto (KEK 1979); R.N.~Mohapatra and
G.~Senjanovi{\'c}, Phys. Rev. Lett. {\bf 44}, 912 (1980);
R.N.~Mohapatra and G.~Senjanovi\'c, Phys. Rev. {\bf D23},165
(1981).
%\cite{Bando:1993ma}
\bibitem{Bando:1993ma} 
  M.~Bando, N.~Maekawa, H.~Nakano and J.~Sato,
  %``A Handy calculating method of Higgs potential in SUSY model,''
  Mod.\ Phys.\ Lett.\ A {\bf 8}, 2729 (1993).
  %%CITATION = MPLAE,A8,2729;%%

 %\cite{Nagao:2012fw}
\bibitem{Nagao:2012fw}
 H.~Nagao and Y.~Shimizu 
  in preparation.

%\cite{Ding:2011qt}
\bibitem{Ding:2011qt} 
  G.~-J.~Ding,
  %``Tri-Bimaximal Neutrino Mixing and the $T_{13}$ Flavor Symmetry,''
  Nucl.\ Phys.\ B {\bf 853}, 635 (2011)
  [arXiv:1105.5879 [hep-ph]].
  %%CITATION = ARXIV:1105.5879;%%
  
%\cite{Hagedorn:2010th}
\bibitem{Hagedorn:2010th} 
  C.~Hagedorn, S.~F.~King and C.~Luhn,
  %``A SUSY GUT of Flavour with S4 x SU(5) to NLO,''
  JHEP {\bf 1006}, 048 (2010)
  [arXiv:1003.4249 [hep-ph]].
  %%CITATION = ARXIV:1003.4249;%%
  
%\cite{Adulpravitchai:2008yp}
\bibitem{Adulpravitchai:2008yp} 
  A.~Adulpravitchai, A.~Blum and C.~Hagedorn,
  %``A Supersymmetric D4 Model for mu-tau Symmetry,''
  JHEP {\bf 0903}, 046 (2009)
  [arXiv:0812.3799 [hep-ph]].
  %%CITATION = ARXIV:0812.3799;%%  
  

%\cite{Chun:2000jr}
\bibitem{Chun:2000jr} 
  E.~J.~Chun, D.~Comelli and D.~H.~Lyth,
  %``The Abundance of relativistic axions in a flaton model of Peccei-Quinn symmetry,''
  Phys.\ Rev.\ D {\bf 62}, 095013 (2000)
  [hep-ph/0008133].
  %%CITATION = HEP-PH/0008133;%% 
  
%\cite{Morrissey:2005uz}
\bibitem{Morrissey:2005uz} 
  D.~E.~Morrissey and J.~D.~Wells,
  %``The Tension between gauge coupling unification, the Higgs boson mass, and a gauge-breaking origin of the supersymmetric mu-term,''
  Phys.\ Rev.\ D {\bf 74}, 015008 (2006)
  [hep-ph/0512019].
  %%CITATION = HEP-PH/0512019;%%
 
%\cite{Cyburt:2004yc}
\bibitem{Cyburt:2004yc} 
  R.~H.~Cyburt, B.~D.~Fields, K.~A.~Olive and E.~Skillman,
  %``New BBN limits on physics beyond the standard model from He-4,''
  Astropart.\ Phys.\  {\bf 23}, 313 (2005)
  [astro-ph/0408033].
  %%CITATION = ASTRO-PH/0408033;%%
  
%\cite{Komatsu:2010fb}
\bibitem{Komatsu:2010fb} 
  E.~Komatsu {\it et al.}  [WMAP Collaboration],
  %``Seven-Year Wilkinson Microwave Anisotropy Probe (WMAP) Observations: Cosmological Interpretation,''
  Astrophys.\ J.\ Suppl.\  {\bf 192}, 18 (2011)
  [arXiv:1001.4538 [astro-ph.CO]].
  %%CITATION = ARXIV:1001.4538;%%
  
  
\bibitem{Barreiro:1996dx} 
For example, See~~
  T.~Barreiro, E.~J.~Copeland, D.~H.~Lyth and T.~Prokopec,
  %``Some aspects of thermal inflation: The Finite temperature potential and topological defects,''
  Phys.\ Rev.\ D {\bf 54}, 1379 (1996)
  [hep-ph/9602263].
  %%CITATION = HEP-PH/9602263;%%
  
%\cite{Asaka:1999xd}
\bibitem{Asaka:1999xd} 
  T.~Asaka and M.~Kawasaki,
  %``Cosmological moduli problem and thermal inflation models,''
  Phys.\ Rev.\ D {\bf 60}, 123509 (1999)
  [hep-ph/9905467].
  %%CITATION = HEP-PH/9905467;%%
  
%\cite{Nelson:1993nf}
\bibitem{Nelson:1993nf} 
  A.~E.~Nelson and N.~Seiberg,
  %``R symmetry breaking versus supersymmetry breaking,''
  Nucl.\ Phys.\ B {\bf 416}, 46 (1994)
  [hep-ph/9309299].
  %%CITATION = HEP-PH/9309299;%%
 \end{thebibliography}
\end{document}